\documentclass[epj,nopacs]{svjour}
\usepackage{graphicx}

\let\I\i
\def\i{\mathrm{i}}

\def\d{\mathrm{d}}
\def\half{{\textstyle{1\over2}}}
\def\thalf{{\textstyle{3\over2}}}
\def\fhalf{{\textstyle{5\over2}}}
\def\h{{\scriptscriptstyle{1\over2}}}
\def\th{{\scriptscriptstyle{3\over2}}}

\def\vec#1{\mbox{\boldmath$#1$}}
\def\CG#1#2#3#4#5#6{C^{#5#6}_{#1#2#3#4}}

\begin{document}

\title{A chiral quark model for meson electro-production 
in the region of D-wave resonances}

\author{%
B.~Golli \inst{1}\thanks{\email{bojan.golli@ijs.si}}
\and
S.~\v{S}irca \inst{2}\thanks{\email{simon.sirca@fmf.uni-lj.si}}
}

\institute{%
Faculty of Education,
              University of Ljubljana and J.~Stefan Institute,
              1000 Ljubljana, Slovenia
\and
Faculty of Mathematics and Physics,
              University of Ljubljana and J.~Stefan Institute,
              1000 Ljubljana, Slovenia}

\date{\today}

\abstract{%
The meson scattering and electroproduction amplitudes in the D13, 
D33 and D15  partial waves are calculated in a coupled-channel 
formalism incorporating quasi-bound quark-model states, extending 
our previous studies of the P11, P33 and S11 partial waves.
The vertices of the baryon-meson interaction including the
$s$- and $d$-wave pions and $\rho$-mesons, the $s$-wave 
$\eta$-meson, and the $s$- and $p$-wave $\sigma$-mesons
are determined in the Cloudy Bag Model, with some changes
of the parameters to reproduce the widths of the resonances.
The helicity amplitudes and the electroproduction amplitudes
exhibit consistent behavior in all channels but tend 
to be too weak compared to the experiment.
We discuss possible origins of this discrepancy which arises
also in the constituent quark model calculations.
}


\maketitle

\section{Introduction}

In our previous research \cite{EPJ2008,EPJ2009,EPJ2011} we 
have investigated the $P$- and $S$-wave nucleon resonances 
in the intermediate energy region using a coupled-channel 
formalism which provides an unified treatment of 
the scattering and the electroproduction processes.
The formalism incorporates in a consistent way the
quark-model resonance states as excitations
of the quark core supplemented by a cloud of mesons.
The most important conclusion of these studies was
that the main component of the resonance state is indeed 
the single-particle excitation of the quark core as predicted 
by the 
quark model in which 
a single-quark is excited either to the $2s$ state (in the 
case of the $P$-wave resonances), or into the $1p$ state 
(in the case of the $S$-wave).
Excitations of the meson cloud may also 
represent an important component of the excited state.
We have found that while the scattering 
amplitudes can be well reproduced in different models of 
resonances by a modest readjustment of model parameters,
the decisive test of the model is the $Q^2$-behavior of 
the electroproduction amplitudes.
By observing the amplitude at lower $Q^2$ and 
intermediate $Q^2$ it may be possible to disentangle 
the contribution of the pion cloud, which dominates at 
the periphery, from the contribution of the quark core.
In general, we have found that the meson cloud plays
an important role, in particular in the EM processes
that are sensitive to the long-range behavior of
the resonance wave-function.
The most evident 
examples
are the dominance of the pion cloud in the quadrupole 
excitation of the $\Delta$(1232) \cite{plb96}
and the zero crossing of the helicity amplitude in the 
$N$(1440) \cite{EPJ2009}.

In the present approach we apply the method to the low-lying
$D$-wave resonances.
Our aim is to check whether the quark core excitation is 
the principal mechanism for the resonance formation
also in this partial wave and, secondly, to study effects
of the meson cloud in order to check whether similar 
effects that were identified in $S$- and $P$-waves
are also presented in the $D$-wave resonances.
In addition, our approach gives an opportunity to study
the two-meson decays as a supplementary method to 
investigate the underlying resonance dynamics.

Experimentally, helicity amplitudes and electroproduc\-tion multipoles
have been extracted from the measured quantities (cross-sections
and polarization observables) in single- and double-pion
electroproduction experiments at Jefferson Lab, Mainz, and Bonn
(see \cite{innavolker12,innareview13} for a review).
In the analysis of single-pion data, unitary isobar models like MAID
\cite{MAID2007,MAID2008,MAID2011} and UIM \cite{Inna09} have been used,
as well as dispersion relations approaches \cite{Inna09}; the two-pion
channels have been analyzed in the JLab-MSU model \cite{mokeev12}.
The multipole amplitudes are also obtained in partial-wave
analyses ({\em e.g.\/} SAID \cite{Arndt06,SAID}, Bonn-Gatchina \cite{BoGa},
Gie\ss en \cite{penner02}, Zagreb \cite{Zagreb10}, and Kent State
\cite{Manley92}), all of which rest on different assumptions
and technical details that in many instances render mutually
inconsistent results.

The low-lying $D$-wave resonances have been explored by systematic
studies of photoproduction and pion-in\-duced production of non-strange
and strange mesons in chiral unitary approaches \cite{Doring07,Doring06},
dynamical coupled-channel approaches like the J\"ulich 2012 model
\cite{ronchen13,julich2012}, and the model developed by EBAC at JLab
\cite{kamano13,julia08,julia09,Matsuyama07}.  
In the quark model only helicity amplitudes have been calculated 
\cite{close72,warns90,capstick95,aiello98,merten02,Santopinto12,Metsch13}.

In the next section we briefly review our method.
We introduce a more general approach to the treatment of the decay 
into an unstable intermediate baryon and a meson in which the 
intermediate baryon decays into two or more channels.
In sect.~3 the quark structure of the considered resonances is 
specified and the parameters of the underlying quark model are 
discussed. 
In sect.~4 we display the results for the scattering amplitudes in 
the considered partial wave and present our prediction for the 
widths  and the branching fractions for the $N$(1520)D13, 
$N$(1700)D13, $N$(1675)D15 and $\Delta$(1700)D33 resonances.
In sect.~5 we discuss the results for the transverse helicity 
amplitudes $A_{1/2}$ and  $A_{3/2}$  for the $N(1520)$, 
$N(1675)$ and $\Delta(1700)$ resonances.
In sect.~6 we review the results for pion photoproduction, 
separately for the $E_{2-}$ and $M_{2-}$  ($M_{2+}$)  amplitudes.
The last section contains some concluding remarks.

\section{\label{sec:basics}Basics of the coupled-channel approach}

In our previous work \cite{EPJ2008} we have developed a method
in which the quasi-bound quark-model states are incorporated 
in the channel states obeying proper asymptotic behavior.
We have shown that such states can be cast in the form
\begin{eqnarray}
|\Psi^{MB}_{JI}\rangle &=& \mathcal{N}_{MB}\left\{
    [a^\dagger(k_M)|\widetilde{\Psi}_B\rangle]^{JI} 
+
 \sum_{\mathcal{R}}c_{\mathcal{R}}^{MB}|\Phi_{\mathcal{R}}\rangle
\right. \nonumber\\ && \left. 
+ \sum_{M'B'}
   \int {\d k\>
       \chi^{M'B'\,MB}(k,k_M)\over\omega_k+E_{B'}(k)-W}\,
      [a^\dagger(k)|\widetilde{\Psi}_{B'}\rangle]^{JI}\right\}\,,
\nonumber\\
\label{PsiH}
\end{eqnarray}
where
the first term represents the free meson ($\pi$, $\eta$, $\rho$, 
$K$, $\dots$) and the baryon ($N$, $\Delta$, $\Lambda, \ldots$) 
and defines the channel, the next term is the sum over 
{\em bare\/} three-quark states $\Phi_{\mathcal{R}}$ involving 
different excitations of the quark core, while the third term 
describes meson clouds around different isobars.
Here $W$ is the invariant energy, $J$ and $I$ are the angular 
momentum and isospin of the meson-baryon system, $\omega_{M}$ 
and $k_{M}$ are the energy and momentum of the incoming (outgoing) 
meson, $\widetilde{\Psi}_{B}$ is a properly normalized 
baryon state (see appendix~\ref{weights}) and  $E_B$ is its energy.
The normalization factor is
$\mathcal{N}_{MB} = \sqrt{\omega_{M} E_{B} / (k_{M} W)}$.
The integration over meson momenta is defined in the principal 
value sense.
Considering chiral quark models in which mesons couple linearly 
to the quark core, the $K$~matrix element between the meson-baryon 
channels (labeled by $MB$ and $M'B'$, respectively) can be written 
in the form
\begin{equation}
 K_{M'B'\,MB}^{JI} =  -\pi\mathcal{N}_{M'B'}
   \langle\Psi^{MB}_{JI}||V_{M'}(k)||\widetilde{\Psi}_{B'}\rangle\,,
\label{defK}
\end{equation}
where $V_{M'}(k)$ stands for the quark-meson vertex of the
underlying quark model and $\widetilde{\Psi}_{B'}$ is the
baryon state in the $M'B'$ channel.

The meson amplitudes $\chi^{M'B'\,MB}(k,k_M)$ are proportional to 
the (half) off-shell matrix elements of the $K$ matrix and
are determined by solving  an equation of the Lippmann-Schwin\-ger type.
The resulting matrix elements of the $K$ matrix take the form
\begin{eqnarray}
   K_{M'B'\,MB}(k,k_M) 
     &=& -\sum_{\mathcal{R}}
     {{\cal V}^M_{B\mathcal{R}}(k_M){\cal V}^{M'}_{B'\mathcal{R}}(k)
      \over Z_{\mathcal{R}}(W) (W-W_{\mathcal{R}})}
\nonumber\\
         && + K^{\mathrm{bkg}}_{M'B'\,MB}(k,k_M)\,,
\label{sol4K}
\end{eqnarray}
where the first term represents the contribution of various 
resonances, while $K^{\mathrm{bkg}}_{M'B'\,MB}(k,k_M)$ 
originates in the non-resonant background processes. 
Here ${\mathcal{V}}^M_{B\mathcal{R}}$ is the dressed 
matrix element of the quark-meson interaction between the 
resonance state and the baryon state in the channel $MB$, 
and $Z_{\mathcal{R}}$ is the wave-function normalization. 
The physical resonance state $\mathcal{R}$ is a superposition 
of the dressed states built around the bare three-quark states 
$\Phi_{\mathcal{R}'}$.
The $T$ matrix is finally obtained by solving the Heitler equation
\begin{equation}
T_{MB\,M'B'} = K_{MB\,M'B'} 
  + \i \sum_{M''B''} T_{MB\,M''B''}K_{M''B''\,M'B'}\,.
\label{Heitler}
\end{equation}

In our approach we make the usual assumption that the two-pion 
decay proceeds either through an  unstable meson ($\rho$-meson, 
$\sigma$-meson, \ldots) or through a baryon resonance 
($\Delta(1232)$, $N^*(1440)$, \ldots).
In such a case, the channel depends either on the invariant mass 
$M_B$ of the $M'' B''$ subsystem into which the resonance decays,
or the invariant mass of the mesons (normally two pions)
of the outgoing unstable meson ($\sigma$ or $\rho$).
The unstable-baryon state is normalized as 
$\langle\widetilde{\Psi}_B(M_B')|
\widetilde{\Psi}_B(M_B)\rangle=\delta(M_B'-M_B)$,
where $M_B$ is the invariant mass of the meson-baryon subsystem.
In such cases, the Heitler equation implies also the
summation (integration) over the invariant masses of either
the baryon-meson or the two-mesons subsystems.
The equation can be simplified by noting that close to 
the resonance the dependence on the invariant mass can be 
expressed in terms of a weight function corresponding to 
a specific decay of the resonance. 
Since in this work we consider the processes at higher energies 
which involve a decay of the intermediate resonance into two
or more channels --- a situation not treated in our previous 
work --- we give details of the construction of the 
orthonormal basis states $\widetilde{\Psi}_B(M_B)$ and
the corresponding weight functions in appendix~\ref{weights}.

Considering meson electroproduction, 
the $T$ matrix for $\gamma^\ast N \to MB$ satisfies
\begin{equation}
  T_{MB\,\gamma N} = K_{MB\,\gamma N} 
                + \i \sum_{M'B'} T_{MB\,M'B'}K_{M'B'\,\gamma N}\,.
\label{Heitler4gamma}
\end{equation}
In the vicinity of a resonance (${\cal R}$) we 
split the $K$ matrix into the ``resonant'' part and
the background which includes also all possible other
resonances in the considered partial wave:
\begin{eqnarray}
   K_{MB\,\gamma N} &=& 
 -{{\cal V}^M_{B\mathcal{R}}{\cal V}^{\gamma}_{N\mathcal{R}}
      \over Z_{\mathcal{R}}(W) (W-W_{\mathcal{R}})}
\nonumber\\
  &&-\sum_{\mathcal{R'}\ne \mathcal{R}}
     {{\cal V}^M_{B\mathcal{R'}}{\cal V}^{\gamma}_{N\mathcal{R'}}
      \over Z_{\mathcal{R'}}(W) (W-W_{\mathcal{R'}})}
       + {B}^{\mathrm{bkg}}_{MB\,\gamma N}\,.
\end{eqnarray}
From (\ref{sol4K}) it follows that the first term can
be written in the form
\begin{equation}
    {{\cal V}^M_{B\mathcal{R}}{\cal V}^{\gamma}_{N\mathcal{R}}
      \over Z_{\mathcal{R}}(W) (W-W_{\mathcal{R}})}
= 
  \left(K_{MB\,\pi N} - K_{MB\,\pi N}^{\mathrm{bkg}}\right)
   {{\cal V}^{\gamma}_{N\mathcal{R}}\over{\cal V}^\pi_{N\mathcal{R}}}
\end{equation}
so that (\ref{Heitler4gamma}) takes the form
\begin{eqnarray}
T_{MB\,\gamma N} &=& 
    {{\cal V}^{\gamma}_{N\mathcal{R}}\over{\cal V}^\pi_{N\mathcal{R}}}\;
      T_{MB\,\pi N}
  +  T^{\mathrm{bkg}}_{MB\,N\gamma}
\nonumber\\
  &\equiv& T^{\mathrm{res}}_{MB\,\gamma N} +  T^{\mathrm{bkg}}_{MB\,\gamma N}\,,
\label{Tgamma}
\end{eqnarray}
which means that the $T$ matrix for electroproduction can be
split into the resonant part and the background part;
the latter is the solution of the Heitler equation with
the ``background'' $K$ matrix defined as
\begin{eqnarray*}
   K^{\mathrm{bkg}}_{MB\,\gamma N} &=& - K_{MB\,\pi N}^{\mathrm{bkg}} 
  \;{{\cal V}^{\gamma}_{N\mathcal{R}}\over{\cal V}^\pi_{N\mathcal{R}}}
\nonumber\\
  &&-\sum_{\mathcal{R'}\ne \mathcal{R}}
     {{\cal V}^M_{B\mathcal{R'}}{\cal V}^{\gamma}_{N\mathcal{R'}}
      \over Z_{\mathcal{R'}}(W) (W-W_{\mathcal{R'}})}
       + {B}^{\mathrm{bkg}}_{MB\,\gamma N}\,.
\end{eqnarray*}
Note that ${\cal V}^{\gamma}_{N\mathcal{R}}(k_\gamma)$ is proportional
to the helicity amplitudes, while the strong amplitude
${\cal V}^M_{B\mathcal{R}}(k_M)$ to $\zeta\sqrt{\Gamma_{MB}}$, 
where $\zeta$ is the sign of the meson decay amplitude.

\section{\label{sec:quark}The D-wave resonances in the quark model}

In the quark model, the negative parity $D$-wave resonances are 
described by a single-quark $p$-wave ($l=1$) orbital excitation.
The two   D13 (flavor octet, $J=\thalf$) resonances are 
superpositions of the spin doublet ($S=\half$) and 
quadruplet ($S=\thalf$) configurations \cite{Isgur77}.
We use the $j$--$j$ coupling scheme \cite{Myhrer84b} in which the 
resonances take the following forms:
\begin{eqnarray}
N(1520) &=& 
  -\sin\vartheta_d|^4{\bf 8}_{3/2}\rangle 
   + \cos\vartheta_d|^2{\bf 8}_{3/2}\rangle 
\nonumber\\  
&&\kern-52pt =  c^l_{S}|(1s)^21p_{3/2}\rangle_{MS} + 
      c^l_{A}|(1s)^21p_{3/2}\rangle_{MA} +
      c^l_{P}|(1s)^21p_{1/2}\rangle \,,
\nonumber\\
\label{N1520}\\
N(1700) &=& 
    \cos\vartheta_d|^4{\bf 8}_{3/2}\rangle 
  + \sin\vartheta_d|^2{\bf 8}_{3/2}\rangle 
\nonumber\\
&&\kern-52pt =   c^u_{S}|(1s)^21p_{3/2}\rangle_{MS} + 
      c^u_{A}|(1s)^21p_{3/2}\rangle_{MA} +
      c^u_{P}|(1s)^21p_{1/2}\rangle\,.
\nonumber\\
\label{N1700}
\end{eqnarray}
Here $1p_{1/2}$ and  $1p_{3/2}$ denote the single-quark states with 
$j=\half$ and $j=\thalf$, respectively, and $MS$ and $MA$ denote the 
mixed symmetric and the mixed antisymmetric spatial representation.
The coefficients are given as
\begin{eqnarray}
c_{S}  &=& {2\over3}\,
      \left\{{\sin\vartheta_d\atop -\cos\vartheta_d}\right\}
    + \sqrt{5\over18}\,
      \left\{{\cos\vartheta_d\atop\sin\vartheta_d}\right\},
\nonumber\\
c_{A} &=& -{\sqrt2\over2}\,
         \left\{{\cos\vartheta_d\atop\sin\vartheta_d }\right\},
\nonumber\\
c_{P} &=& {\sqrt5\over3}\,
  \left\{{-\sin\vartheta_d\atop\cos\vartheta_d }\right\} 
  + {\sqrt2\over3}\,
   \left\{{\cos\vartheta_d\atop\sin\vartheta_d}\right\}\,,
\nonumber\\
\label{cSAP}
\end{eqnarray}
where the upper values in $\{\}$ refer to the $N(1520)$ resonance.
The constituent quark model calculations (see {\em e.g.\/} 
\cite{Isgur77}) as well as the calculation in the MIT bag model with 
hyperfine interaction \cite{Myhrer84b,deGrand76b} predict that the 
$^4{\bf 8}$ configuration is some 150~MeV higher than the $^2{\bf 8}$, 
suggesting a small value of the mixing angle $\vartheta_d$.
The small value agrees with the quark-model analysis of the 
$\pi N$ decay of the resonances (see {\em e.g.\/} \cite{Hey75}) 
which predicts a nearly negligible decay amplitude of the
$N$(1700) into the $\pi N$ channel.
Let us note that in the dynamical coupled-channel approaches
\cite{ronchen13,oset13} the $N(1700)$ resonance appears to be
dynamically generated, with strong couplings to $\rho N$
and $K^*\Lambda$ channels.

The D33 resonance (flavor decuplet) has $S=\half$, while the 
D15  resonance (octet, $J=\fhalf$) has $S=\thalf$, thus
\begin{eqnarray}
\Delta(1700)D33 &=& |^2{\bf 10}_{3/2}\rangle 
\nonumber\\
  &=&   {\sqrt5\over3} \vert (1s)^21p_{3/2} \rangle
    - {2\over3}\vert (1s)^21p_{1/2} \rangle\,,
\label{D1700}
\\
N(1675)D15 &=& |^4{\bf 8}_{5/2}\rangle 
  =   \vert (1s)^21p_{3/2} \rangle\,.
\label{N1675}
\end{eqnarray}
We shall not consider the D35 partial wave
since the experimental data in this case are 
too scarce. 

The underlying chiral quark model in our calculations 
in the P11, P33 and S11 partial waves has been
the Cloudy Bag Model \cite{CBM}.
In these studies we kept its parameters fixed 
to the popular values used in the calculations of nucleon 
properties, {\em i.e.\/} the bag radius 
$R=0.83$~fm, which determines the range of the quark-pion interaction 
corresponding to the cut-off $\Lambda\approx 550$~MeV/c, and 
$f_\pi=76$~MeV, which reproduces the experimental
value of the $\pi NN$ coupling constant.

The $l=2$ pions couple only to $j=3/2$ quarks; the corresponding
interaction in the Cloudy Bag Model takes the form
\begin{eqnarray}
V^\pi_{2mt}(k) &=& {1\over2f_\pi}\sqrt{\omega_{p_{3/2}}\omega_s\over
     (\omega_{p_{3/2}}-2)(\omega_s-1)}\,
     {\sqrt2\over2\pi}\,{k^2\over\sqrt{\omega_k}}\,{j_2(kR)\over kR}
\nonumber\\
     &&\times \sum_{i=1}^3 \tau_t(i)\,\Sigma^{[\h\th]}_{2m}(i)\,,
\label{Vpid}
\end{eqnarray}
where $\omega_s=2.043$, $\omega_{p_{3/2}}=3.204$, and
$$
\Sigma_{2m}^{[\h\th]}
    = \kern-3pt\sum_{m_sm_j}\kern-3pt\CG{\th}{m_j}{2}{m}{\h}{m_s}
      |sm_s\rangle\langle p_{3/2}m_j|\,.
$$

In our previous work \cite{EPJ2011} we have introduced the 
quark-meson coupling of other members of the SU(3) meson octet, 
as well as the coupling of the $\sigma$- and $\rho$-mesons.
In the formulas given for the $\rho$-quark couplings we have
considered only the mesons with transverse polarization, which 
is justified in the case of positive-parity resonances.
A more complete treatment including the construction of $\rho N$ 
channels with good spin of the $\rho N$ system along with the discussion 
on the $\sigma$-quark coupling is given in appendix~\ref{rhoN}.

Assuming SU(3) flavor symmetry all coupling constants of the
meson octet are fixed by the value of the $\pi NN$ coupling constant.
The $\sigma$ and $\rho$ couplings are free parameters in principle.
Since the results turn out to depend only weakly on these values,
and because the data for the corresponding channels are rather uncertain,
we assume the same value for the $\sigma$ coupling as for the one used
in the P11 partial wave, while for the $\rho$-meson we assume simply
$f_\rho = f_\pi$.

In order to reproduce the decay widths, as described in the following
section, we had to increase the quark-model values for the $d$-wave
quark-meson couplings and either increase or decrease the $s$-wave
couplings, while preserving the $p$-wave couplings.
We note that the wave-function and vertex renormalization through
meson loops does not solve the problem of the widths as it has been
the case in the Delta(1232) and the N(1440) \cite{EPJ2008}.
In fact, there is a tendency that they even diminish with respect
to their bare values by $10\,\%$ to $20\,\%$.
The $p$-wave couplings are not necessarily the bare values; e.g.,
the $\pi N\Delta$ coupling is the dressed one, as determined
in our calculation in the P33 case.

We use a single renormalization factor for all $d$-wave couplings
and another one for all $s$-wave couplings in the given partial
wave (with one exception in the D15 wave), thus keeping the number
of free parameters small.  The fact that only a single factor is needed
would follow naturally if the behavior of the quark wave-functions
were improved.

The predictive power of the quark model can be judged upon
two considerations: by assessing the corrections need\-ed
to reproduce the pion-nucleon scattering amplitudes, and by
comparing the helicity or total photo-production amplitudes to data.
It is crucial to note that the latter receives input from the former,
hence the strong part needs to be controlled well before being
able to assess the quality of the electromagnetic contribution.

In addition to the renormalization factors, the adjus\-table parameters
are the positions of the $K$-matrix poles of the resonances.  Typical
differences between bare and pole masses are: 200~MeV (for D13),
240~MeV (D15), and 500~MeV (D33).  These values are generally
smaller compared to those in the calculations in dynamical models
(see e.g.~\cite{ronchen13}) as a consequence of the smaller cut-off
(i.e.~larger bag radius) used in our calculation.

\section{\label{sec:scattering}The scattering amplitudes}

The behavior of the scattering amplitudes in the D13 partial 
wave is governed by a subtle interplay of the elastic and  
--- primarily --- the $s$-wave $\pi\Delta$ channel.
As we have mentioned in the previous section, the lower 
$N(1520)$ resonance is predominantly the $^2{\bf 8}$ configuration.
It is strongly coupled to the $\pi N$ channel and moderately 
to the  $s$-wave $\pi\Delta$ channel.
The upper $N(1700)$ resonance is then mainly the $^4{\bf 8}$ 
configuration; its quark model coupling to the $\pi N$ channel 
is a factor of $3\sqrt{5}$ weaker than in the  $^2{\bf 8}$ 
configuration, but  stronger in the case of the $\pi\Delta$ channel.
The corresponding ratio in the  $\pi\Delta$ channel is $\sqrt{5/2}$ 
for the  $s$-wave and $-\sqrt{8/5}$ for the $d$-wave pions.
As already noted by Hey et al. \cite{Hey75} this explains
qualitatively the observed behavior of the decay amplitudes.

\begin{figure}[h!]
\begin{center}
\includegraphics[width=88mm]{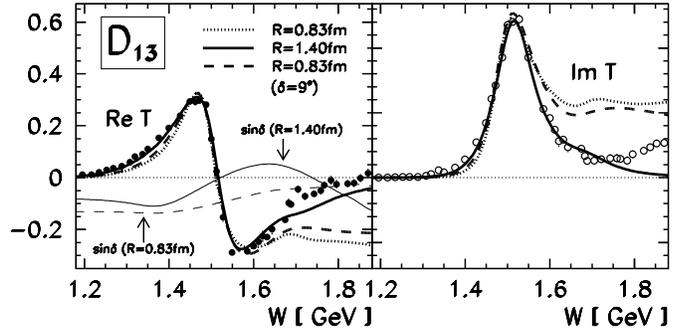}
\end{center}
\caption{The real and imaginary part of the elastic scattering
$T$ matrix for the D13 partial wave for three choices of parameters:
bag radius $R=0.83\,\mathrm{fm}$ (thick dotted line), 
$R=1.40\,\mathrm{fm}$ (thick full line), both with $W$-dependent
$\vartheta_\delta$, and $R=0.83\,\mathrm{fm}$ with fixed
$\vartheta_\delta = 9^\circ$ (thick dashed line).
The data points are from the SAID $\pi N\to\pi N$ partial-wave 
analysis \cite{Arndt06,SAID}.
Also shown is the sine of the mixing angle between spin $\half$
and $\thalf$ configurations defined in (\ref{N1520}) and
(\ref{N1700}).}
\label{fig:TD13}
\end{figure}

In our coupled-channel calculation we have included in addition
to the elastic and the $s$- and $d$-wave $\pi\Delta$ channels also 
the $p$-wave $\sigma N$ channel, the $d$-wave $\pi N(1440)$ and 
$\eta N$ channels, as well as the $s$- and $d$-wave $\rho N$ channels. 
For our standard choice of the model parameters 
discussed in the previous section the calculated elastic amplitude 
turns out to be too weak, while the $s$-wave $\pi\Delta$ decay 
amplitude is overestimated.
This effect is further enhanced if we allow for the mixing
of the two resonances  through $s$- and $d$-wave pion loops
with the intermediate $\Delta$ and the nucleon.
The elastic width of the lower resonance is reproduced only if we
increase the $d$-wave pion coupling to the $\pi N$ channel
by 50~\% and reduce the $s$-wave coupling to the $\pi\Delta$
channel by almost 40~\%.

\begin{figure}[h!]
\begin{center}
\includegraphics[width=80mm]{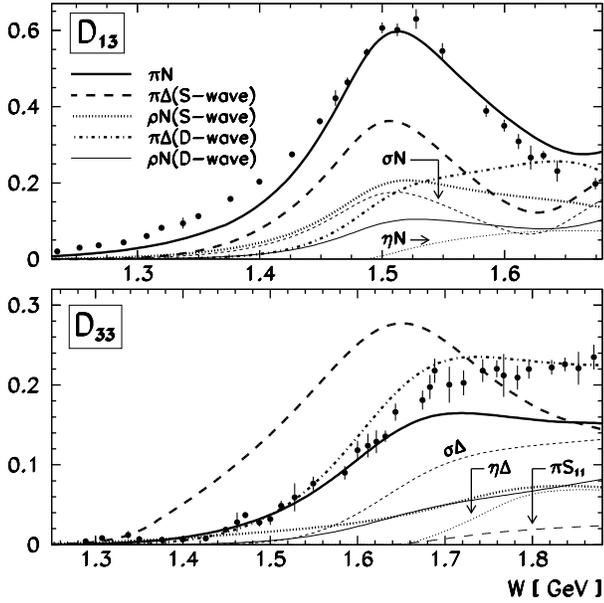}
\end{center}
\caption{The absolute values $|T_{MB\,\pi N}|$ of the amplitudes 
for the elastic and the dominant inelastic channels in the 
D13 partial wave (top panel) and in the D33 partial wave 
(bottom panel).
The $\sigma$ meson and the pion in $\pi S_{11}$ channel are 
in the relative $p$-wave and the $\eta$ meson in the $s$-wave.}
\label{fig:TD13D33ch}
\end{figure}

The agreement with the data improves
considerably if we  increase the bag radius to 1.4~fm.
If we increase simultaneously the couplings of the $d$-wave pions 
by 35~\% and reduce the $s$-wave $\pi\Delta$ coupling by 15~\% 
with respect to their quark-model values, we obtain an almost 
perfect agreement with experiment, as shown in fig.~\ref{fig:TD13}.
As we discuss in the next section, the results for the magnetic 
quadrupole excitation amplitude also favor larger bag radii, 
however, the dominant electric dipole excitation diminishes
considerably in such a case.
The effect on the choice of the mixing angle $\vartheta_d$
introduced in (\ref{N1520}) is shown for a fixed value 
$\vartheta_d=9^\circ$ (close to the value suggested by \cite{Hey75} 
and \cite{Isgur77}) and the value calculated through the pion 
loops which introduces the energy dependent mixing.

We therefore keep the radius at $R=1$~fm which 
still yields consistent values for the ground-state properties.
At this value, the $d$-wave coupling strength has to be 
increased by 43~\% with respect to its quark-model value 
in order to reproduce the experimental width of the resonance, 
$\Gamma=115$~MeV.
The inelastic channels are still dominated by the $s$-wave 
$\pi\Delta$ channel (with the $s$-wave coupling reduced 
to 58~\% of its quark-model value) as seen in 
table~\ref{table:fraction} and fig.~\ref{fig:TD13D33ch}.
As mentioned in the previous section, we use 
$\tilde{f}_\rho=f_\pi$ for the parameter appearing in 
the $\rho$-quark coupling (see appendix~\ref{rhoN}).
This is somewhat stronger with respect to the value that 
would follow from the conventional values for the 
$\rho NN$ coupling; still the branching fraction for 
the $s$-wave $\rho N$ channel is below the PDG value.
The branching fraction for the $\eta N$ channel at 1520~MeV
(not displayed in table~\ref{table:fraction}) is 0.10~\%,
close to the value found in \cite{Tiator99,SAID,ronchen13}
but smaller compared to that in \cite{penner02}.

\begin{figure}[h!]
\begin{center}
\includegraphics[width=88mm]{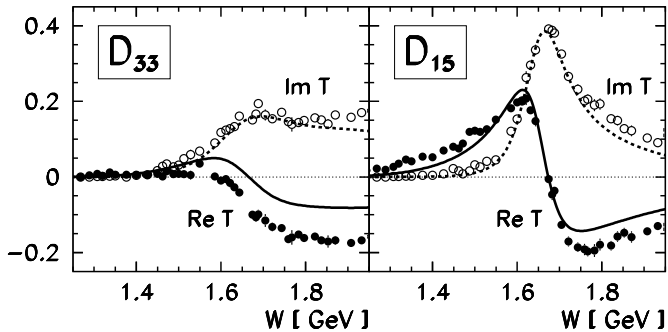}
\end{center}
\caption{The real and imaginary part of the elastic scattering
$T$ matrix for the D33 (left) and D15 (right) partial wave.
The data points are the same as in fig.~\ref{fig:TD13}.}
\label{fig:TD33D15}
\end{figure}

The scattering amplitudes in the D33 partial wave are well 
reproduced for our standard choice of the bag radius 
($R=0.83$~fm), provided the $d$-wave pion coupling strength 
is multiplied by a factor 2.4 which gives the total width 
of 288~MeV (fig.~\ref{fig:TD33D15}).
In order to be able to reproduce the almost flat behavior 
of the elastic amplitude above the resonance we have included 
up to 13 inelastic channels: the $s$- and $d$-wave 
$\pi\Delta$ and $\rho N$ channels, the $d$-wave $\pi N(1440)$ 
channel, the $p$-wave $\pi N(1535)\to \pi\pi N$, 
$\pi N(1535)\to \pi\eta N$, $\pi N(1650)$, $\sigma\Delta$ 
and $\eta\Delta$ channels, and the $s$-wave $\pi\Delta(1600)$ 
and $\sigma \Delta(1700)$ channels.
The  results for the major contributions are presented 
in table~\ref{table:fraction} and in fig.~\ref{fig:TD13D33ch}.
Since the width and the branching fractions are evaluated at
the $K$-matrix pole ($W=1680$~MeV) which is relatively low
compared to the threshold, some of the interesting channels,
such as the $\eta\pi N$ channel, are not included in the table.
In fig.~\ref{fig:TD13D33ch} we notice that the contribution
of the $\eta\Delta$ channel becomes sizable only above 
1700~MeV and remains stronger compared to the competitive 
process in which the $\eta\pi N$ final state is reached 
through the $N$(1535) intermediate state.

The resonant contribution to $\pi N\rightarrow \rho N$ turns 
out to be quite small due to a strong cancellation between the 
$\rho$ couplings to $p_{j=1/2}$ and $p_{j=3/2}$ quarks.
The relatively strong amplitude in fig.~\ref{fig:TD13D33ch}
stems from the background $u$-channel process involving
predominantly the D13 intermediate state.

{\renewcommand{\arraystretch}{1.3}
\setlength{\tabcolsep}{3pt}
\begin{table}[h]
\begin{center}
\begin{tabular}{|l|c|c|c|c|c|c|c|}
\hline

\hline
Res. & $\pi N$    & $\pi\Delta$ ($S$) & 
$\pi\Delta$ ($D$) & $\rho N$ & $\sigma $ \\
\hline
$N(1520)$ &   59~\%   &  23~\%    &    5~\%    & 7~\% ($S$)   &   5~\%  \\
PDG       & 55--65~\% & 10--20~\% & 10--15~\%  & $9\pm 1$~\%  & $<8$~\% \\
\hline
$N(1700)$ &   11~\%   & 35~\%     &   26~\%    & 1~\% ($S$)   & 25~\% \\
PDG   & $12\pm 5$~\%  & 10--90~\% & $<20$~\%   & $7\pm 1$~\%  &     \\
\hline
$\Delta(1700)$ & 15~\% &  50~\%    & 29~\%     &   4~\% ($S$) & 4~\% \\
PDG       & 10--20~\%  & 25--50~\% & 5--15~\%  &  5--20~\%    &     \\
\hline
$N(1675)$ &    39~\%   &   -      & 58~\%      & 2~\% ($D$)   &  -  \\
PDG       & 35--45~\%  &   -   & $50\pm 15$~\% & $1\pm 1$~\%  &  -  \\
\hline

\hline
\end{tabular}
\end{center}
\caption{
The branching fractions for $N$(1520)D13, $N$(1700)D13,
$\Delta$(1700)D33 and $N(1675)$D15.   For the first three
resonances only the $s$-wave $\rho N$ values are compared;
$\sigma$ denotes the $\sigma N$ channel for the D13 resonances
and $\sigma\Delta$ for the D33 case.
The PDG values are from \cite{PDG}.}
\label{table:fraction}
\end{table}}

\begin{figure}[h!]
\begin{center}
\includegraphics[width=88mm]{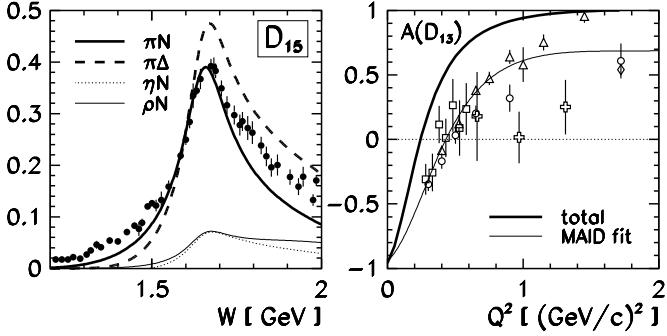}
\end{center}
\caption{The absolute values $|T_{MB\,\pi N}|$ of the 
scattering amplitudes in the D15 partial wave (left)
and the helicity asymmetry $A=(A_{1/2}^2-A_{3/2}^2)/(A_{1/2}^2+A_{3/2}^2)$
for the $N(1520)$D13 (right).}
\label{D15chhelA}
\end{figure}

The scattering amplitudes in the D15 partial waves are 
dominated by the elastic and the $d$-wave $\pi\Delta$ channel
(fig.~\ref{fig:TD33D15} (right) and \ref{D15chhelA} (left)).
We use $R=1$~fm for the bag radius.
Similarly as in the case of the D33 partial wave, the $\pi N$ 
coupling  has to be increased by a factor of $2.25$,
and the $\pi\Delta$, $\eta N$ and $\rho N$ couplings 
by a factor of $1.45$ compared to their quark-model values 
in order to reproduce the experimental width of 150~MeV
and the branching fractions (table~\ref{table:fraction}).
(The branching fraction for the $\eta N$ channel is 1.8~\%.)
In the present model the quarks are excited only to the
$p$-state, so they do not couple to the $\sigma$-meson;
the relatively large fraction of the $\sigma N$ decay 
seen in the experiment may indicate that the excitation
to the $f$-state is important.

\section{\label{sec:helicities}The helicity amplitudes
for the $D$-wave resonances.}

The resonant part of the electroproduction amplitude, proportional
to $M_{MB\gamma N}^\mathrm{res}$ in eq.~(\ref{Tgamma}), reads:
$$
{\mathcal{M}_{MB\gamma N}^\mathrm{res}}  =
\sqrt{\omega_\gamma E_N^\gamma \over \omega_\pi E_N }\,
{\xi\over\pi{\cal V}_{N\mathcal{R}}^\pi}\,
  \langle\widehat{\Psi}_{\mathcal{R}}|{V}_\gamma
                |\Psi_N\rangle\, {T_{MB\,\pi N}} \>,
$$
where $V_\gamma$ is the interaction
of the photon with the electromagnetic current, which contains
quark and pion contributions, and $\xi$ is the spin-isospin factor
depending on the considered multipole and the spin and isospin
of the outgoing hadrons.  The matrix element
$\langle\widehat{\Psi}_{\mathcal{R}}|{V}_\gamma|\Psi_N\rangle$
is the helicity amplitude.  The resonance state
$\widehat{\Psi}_{\mathcal{R}}$ is extracted from the components
in the second and the third term in (\ref{PsiH}) that are
proportional to the resonance pole $(W-W_{\mathcal{R}})^{-1}$;
it involves the bare-quark core and the meson cloud:
\begin{equation}
|\widehat{\Psi}_{\mathcal{R}}\rangle
=   
Z_{\mathcal R}^{-{1\over2}}\left[|{\Phi}_{\mathcal{R}}\rangle
   - \sum_{MB}\int{\d k\quad{\cal{V}}^M_{B{\mathcal R}}(k)
             \over\omega_k+E_B-W}\,
      [a^\dagger(k)|\widetilde{\Psi}_B\rangle]^{JI}
\right].
\label{PsiR}
\end{equation}
Note that the integration is meant in the principal-value sense,
hence the helicity amplitudes are real.
In our model the transverse helicity amplitudes are linear 
combinations of the electric dipole and the magnetic 
quadrupole amplitudes, see {\em e.g.\/} \cite{MAID2007}.
The latter involve only the $1p_\th$ quark state and 
no $1p_\h$.

\begin{figure}[h!]
\begin{center}
\includegraphics[width=88mm]{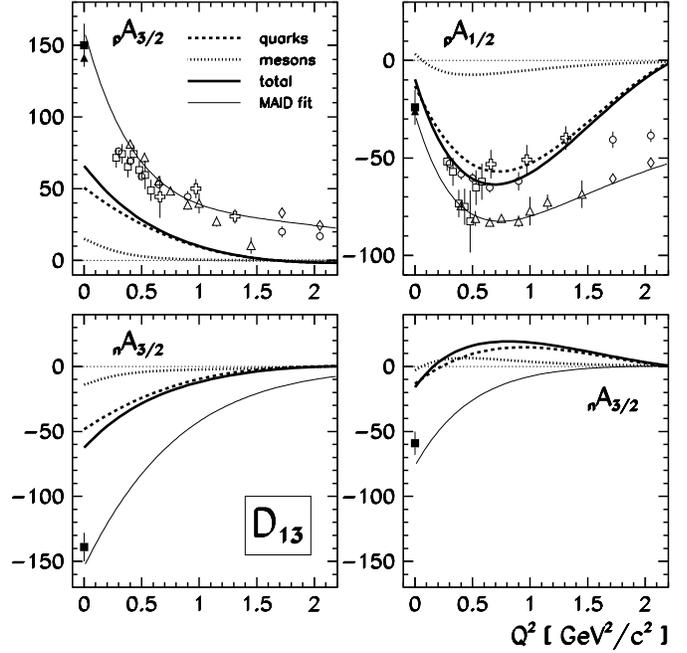}
\end{center}
\caption{Helicity amplitudes for electroexcitation of the D13 
resonance. Top panels: proton target. Bottom panels: neutron target.  
The data points are: the PDG values \cite{PDG} (filled squares), 
pion photoproduction data from CLAS \cite{dugger09} (filled 
triangles), average of dispersion-relation analyses and 
unitary-isobar model of \cite{Inna09} (empty circles), JLab-MSU 
analysis of two-pion electroproduction at CLAS \cite{mokeev12}
(empty squares), MAID2007 analysis \cite{MAID2007} based on 
cross-sections of refs.~\cite{Joo02,Lav04} (empty triangles), 
MAID2008 reanalysis \cite{MAID2008} based on cross-sections of 
ref.~\cite{Par02} (empty diamonds) and JLab two-pion analysis 
\cite{innareview13} (empty crosses).}
\label{fig:AD13}
\end{figure}

\begin{figure}[h!]
\begin{center}
\includegraphics[width=88mm]{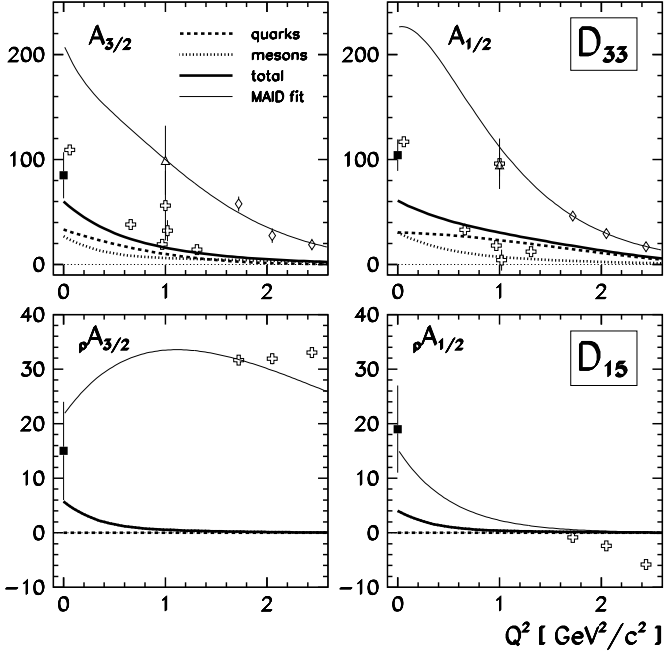}
\end{center}
\caption{Helicity amplitudes for electroexcitation of the D33
(top panels) and D15 resonance (bottom panels).  Notation for
data points as in fig.~\ref{fig:AD13}.}
\label{AD33D15}
\end{figure}

In general, the amplitudes for the $D$-wave resonances
presented in figs.~\ref{fig:AD13} and \ref{AD33D15}
are underestimated with respect to the amplitudes
extracted from experiments in various analyses.
This is a similar situation  as in the
case of the $D$-wave scattering amplitudes
and may again indicate that the description of the peripheral part 
of the resonance wave-function is inadequate.
However, such a conclusion holds also for calculations
in other quark models; the constituent quark model
\cite{aiello98,Santopinto12,Metsch13} predicts a very similar
behavior as our model for the helicity amplitudes 
of the resonances in the D13, D33 and D15 partial wave.
On the other hand, the helicity asymmetry for the $N(1520)$
(fig.~\ref{D15chhelA} (right)) confirms the changeover
to the helicity-$1/2$ dominance at higher $Q^2$, and follows
the general trend in the quark model first predicted by \cite{close72}.

In our calculation, the effect of the  meson cloud, {\em i.e.\/} 
processes in which the photon couples directly to the pion as well 
as the vertex corrections, are generally relatively strong. 
In the constituent quark model, however, the effects of the pion cloud 
have not been taken into account in the considered partial waves;
but as shown in \cite{Ramalho08,Ramalho10} for the P33 partial-wave,
these effects can be strong in the constituent quark models too.
We can therefore conjecture that using a more elaborate model to 
describe the quark core and the meson cloud might eventually bring 
the calculated amplitudes in agreement with the experiment ---
considering also the large scatter of experimental values.
The meson cloud contribution to the D13 helicity amplitudes  
calculated in the Dynamical Coupled Channel Approach \cite{julia08} 
shows even a stronger contribution than in the present calculation,
which could be attributed to the inclusion of the $\gamma\rho\pi$ 
and $\gamma\omega\pi$ vertices that are absent in our approach.
Similarly, the importance of the meson-cloud contribution found
in our approach in the case of the $\Delta$(1700)D33 resonance is
in line with the chiral unitary approach \cite{Doring07},
in which the entire radiative width comes from the meson cloud.

In the case of the $N$(1675)D15 resonance 
we observe a strong deviation of the calculated 
$A_\th$ and $A_\h$ amplitudes at larger $Q^2$ compared 
to the experiment \cite{MAID2011}.
As we have mentioned, in the present approach we assume 
only $p$-wave excitation of the quark core and the $s$- and $d$-wave 
excitation of the meson cloud, in which case only the magnetic 
quadrupole excitation contributes and the helicity 
amplitudes are simply related by $A_\th=\sqrt2A_\h$.
Furthermore, the quark contribution to the isoscalar proton 
amplitude cancels the isovector one, which is the main
reason that the amplitudes almost vanish at larger $Q^2$.
In order to reproduce the behavior of the amplitudes as 
extracted from the experiment, a rather strong contribution 
of the $f$-wave quark excitation has to be assumed; 
note that the importance of the $f$-wave excitation has been 
mentioned already in the previous section as a possible 
explanation of the relatively large decay probability into 
the $\sigma N$ channel.

\section{\label{sec:photoproduction}The photoproduction 
amplitudes}

The photoproduction amplitudes consist of the resonant
and the background contribution.
In our approach, the background term originates from the 
pion pole which governs the amplitudes at low energies,
the contribution from the $u$-channel processes and,
in the case of the D13 partial wave, from the contribution 
of the upper resonance.
In the vicinity of a resonance, the amplitudes are dominated 
by the resonant contribution which turn out to be significantly 
underestimated, as could be anticipated from our results for 
the helicity amplitudes in the previous section.

\begin{figure}[h!]
\begin{center}
\includegraphics[width=88mm]{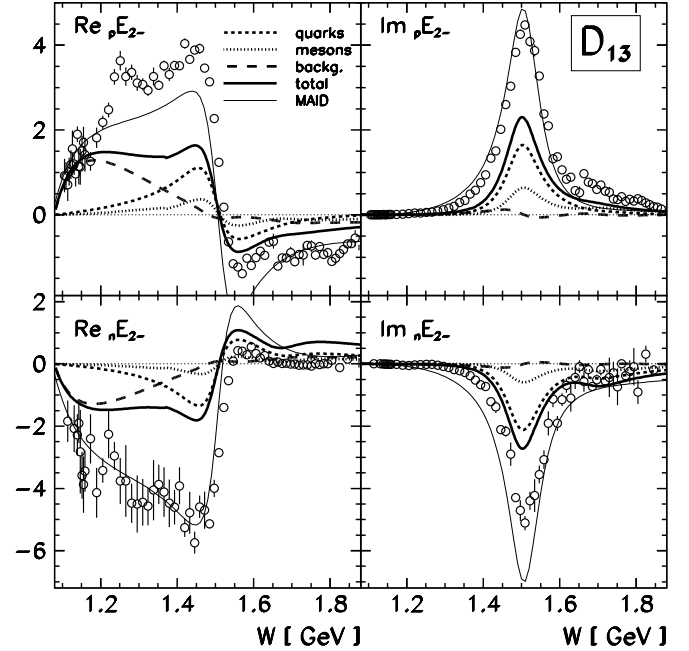}
\end{center}
\caption{The real and imaginary parts of the $E_{2-}$ electroproduction
amplitude (in units of $10^{-3}/m_\pi$) in the D13 partial wave,
for the proton (top panels) and neutron target (bottom panels).
The curves corresponding to our calculation are: resonance quark core
(dashed) and meson cloud (dotted), background (long-dashed), and total
(thick full lines).  The experimental points are taken from
the SAID analysis \cite{SAID}.  The MAID fit is shown by thin full lines.}
\label{fig:E2D13}
\end{figure}

Regarding the magnetic quadrupole contribution, the agreement 
with the experiment can be considerably improved by increasing 
the bag radius to $R\approx 1.4$~fm.
However, the electric dipole contribution becomes even smaller 
in such a case; also, the amplitude drops to zero too quickly
at larger $Q^2$.

\begin{figure}[h!]
\begin{center}
\includegraphics[width=88mm]{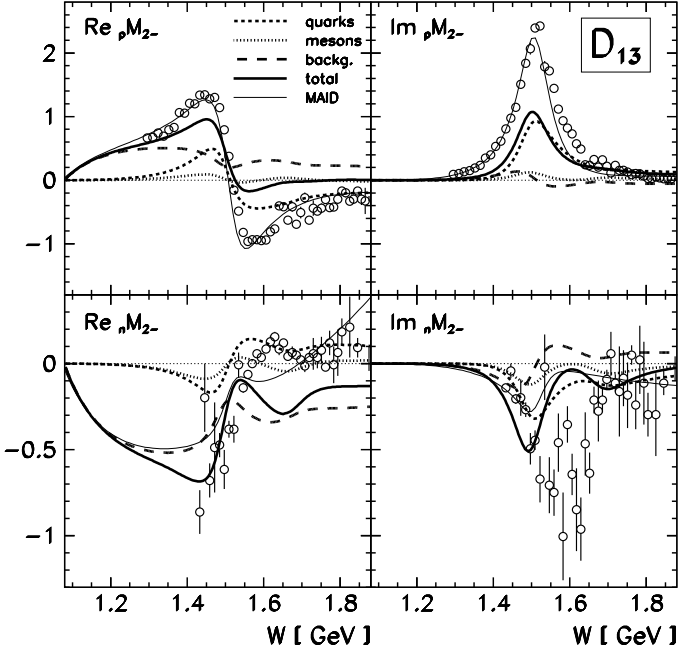}
\end{center}
\caption{The real and imaginary parts of the $M_{2-}$ electroproduction
amplitude for the D13 partial wave.  Notation as in 
fig.~\ref{fig:E2D13}.}
\label{fig:M2D13}
\end{figure}

Although the strength of all amplitudes is underestimated, 
the amplitudes do show a consistent behavior for all multipoles 
and for all partial waves.
In particular, our results for the $E_{2-}$ amplitude in the
D13 partial wave agree with the observation that the amplitudes
for the neutron target are slightly stronger than those for 
the proton, while the $M_{2-}$ amplitude is correctly predicted 
to almost vanish in the neutron case.

In the D33 partial wave we stress the important effects
of the resonance meson cloud which accounts for almost
half of the resonant contribution, bringing the $E_{2-}$
amplitude close to the experiment; in the case of $M_{2-}$
this contribution is, however, still too weak and indicates
an inadequate description of the resonance periphery.

The effect of the meson cloud is not so pronounced in the D15 
partial wave; it represents, however, the sole contribution
in the proton case, since the isoscalar and the isovector
quark contributions cancel.
It is possible that the inclusion of the $f$-quark excitation
is necessary in this case, as suggested from our analysis of 
the inelastic scattering and helicity amplitudes.

\begin{figure}[h!]
\begin{center}
\includegraphics[width=88mm]{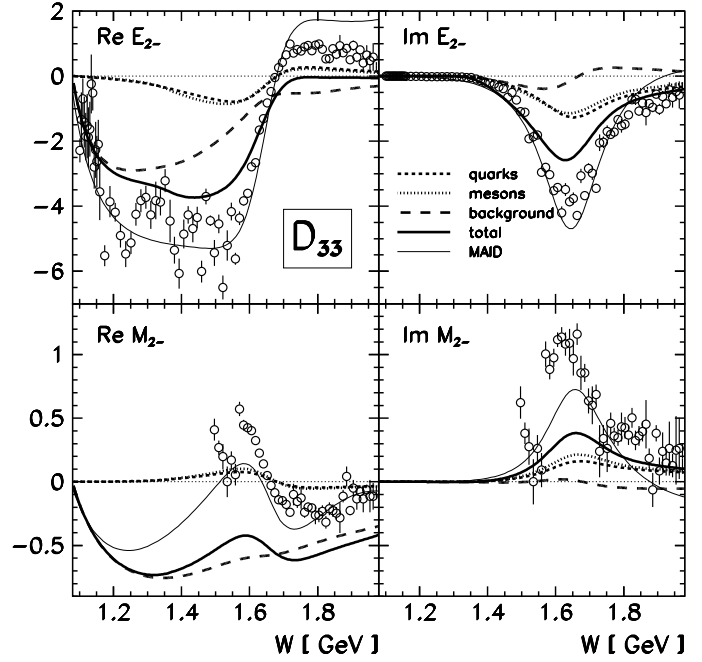}
\end{center}
\caption{The real and imaginary parts of the $E_{2-}$ (top panels)
and $M_{2-}$ (bottom panels) electroproduction amplitudes in
the D33 partial wave.  Notation as in fig.~\ref{fig:E2D13}.}
\label{E2M2D33}
\end{figure}

\begin{figure}[h!]
\begin{center}
\includegraphics[width=88mm]{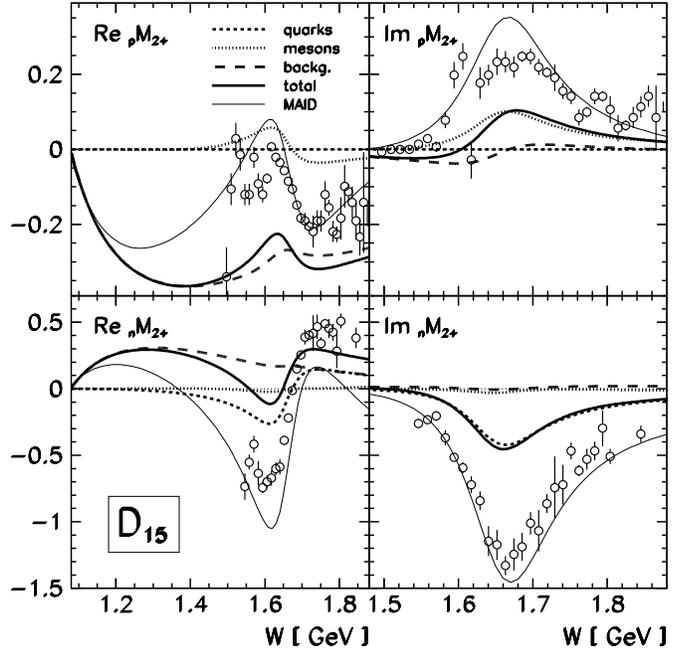}
\end{center}
\caption{The real and imaginary parts of the $M_{2+}$ electroproduction
amplitudes in the D15 partial wave, for the proton (top panels)
and neutron targets (bottom panels).  
Notation as in fig.~\ref{fig:E2D13}.}
\label{M2D15}
\end{figure}

\section{\label{sec:conclusions}Conclusions}

While the results of our model for the scattering and 
electroproduction amplitudes in the case of the P11, P33 
and S11 resonances provide good agreement 
with the experiment, in accordance with the limited scope 
of the rather simple underlying chiral quark model, 
the results for the $D$-wave resonances show a more 
pronounced disagreement, in particular for the prediction 
of the $d$-wave meson coupling to the quark core.
The Cloudy Bag Model sharply cuts the quark wave-functions
at the bag radius, and we cannot expect to be able to describe 
sufficiently well the peripheral region of the resonance
to which the $d$-wave pions are sensitive; the same is
true for the photon electric dipole and magnetic quadrupole
interactions, which both involve $l=2$ photons.
In order to reproduce the observed widths of the 
resonance it was therefore necessary to increase {\it ad hoc\/} the 
quark-model coupling constant by a factor of 1.4 in the 
case of the $N$(1520) and $N$(1700) and even by 2.4 in 
the case of the $\Delta(1700)$ and $N$(1675). 
Furthermore, the model predicts too small helicity 
amplitudes and consequently also the electroproduction 
amplitudes for the considered resonances.
We have not tried to readjust the strength of 
the EM interaction to improve the agreement. 

In order to get an insight into possible origins of the 
disagreement we have increased the bag radius and have been 
able to better reproduce the scattering amplitude in the case 
of the D13 partial wave as well the $M2$ (quad\-ru\-po\-le) amplitude. 
This suggests that a better description of the peripheral
part of the resonance wave-function is needed. 
The increase of radius, however, considerably spoils
the inner part of the wave-function.
Furthermore, our analysis of the scattering and helicity
amplitudes indicates that the higher excitations ({\em i.e.\/} 
the $f$-state) of the quark core has to be included.
Nonetheless, other quark model calculations using more
sophisticated wave-functions obtain similar results for the 
helicity amplitude at least at small and modest $Q^2$.
Our results, in particular for the D33 wave, show that 
the meson cloud effects are important in describing 
the long-range part of the wave-function and may eventually 
bring the results in the ballpark of acceptable values,
which appear to be uncertain at the moment.
In comparison with the P11 and P33 partial waves, which are dominated
solely by $p$-wave mesons, the treatment of the meson cloud effects
is much more sophisticated in the present case, where we encounter 
components of $s$-, $p$- and $d$-wave meson contributions 
of similar strengths.
In fact, we have noticed a sizable cancellation of 
different contributions of the meson cloud, {\em e.g.\/} the 
vertex correction due to pion loops and the contribution 
from the direct coupling of the photon to the pion.
It is therefore possible that in a more elaborate
approach the meson-cloud effects would turn out to be 
stronger and would improve the agreement with experiment.

In spite of the deficiencies discussed above, we conclude
that the overall qualitative agreement with the multipole 
analysis in the D13, D33 and D15 partial waves indicates
that the quark-model explanation of the $D$-wave resonance
as a $p$-wave excitation of the quark core, supplemented
by the meson cloud, is sensible and that no further 
degrees of freedom are needed.

\begin{acknowledgement}
\end{acknowledgement}


\appendix
\section{\label{weights}The weights for the unstable channels}
The PV states (\ref{PsiH}) are normalized as
\begin{equation}
   \langle\Psi^\alpha(W) |
                     \Psi^\beta(W')\rangle
  = \delta(W-W') \left[\delta_{\alpha,\beta} 
  + {\mathbf{K}^2}_{\alpha,\beta}\right]\,.
\label{normPV}
\end{equation}
The PV states are not orthonormal; the orthonormalized states
that enter the definition of the $K$ matrix (\ref{defK})  
are constructed by inverting the norm:
\begin{equation}
  |\widetilde{\Psi}^\alpha(W)\rangle
  = \sum_\beta{\left[\mathbf{1} +
           \mathbf{K}^2\right]^{-1/2}}_{\beta\alpha}
                                      |\Psi^\beta(W)\rangle\,.
\label{orthonorm}
\end{equation}

We now explicitly construct the orthonormal states for the 
interesting region of energies close to a chosen resonance.
Let us first note that in the vicinity of a resonance, 
${\cal R}$, the PV state is dominated by the terms containing 
the pole, {\em i.e.\/} the quasi-bound quark state $\Phi_{\mathcal{R}}$
and the corresponding component in the meson cloud
(\ref{PsiR}):
\begin{eqnarray}
|\Psi^{MB}_{JT}\rangle &=& 
 \mathcal{N}_{MB}\, c_{\mathcal{R}}^{MB}
         |\widehat{\Psi}_{\mathcal{R}}\rangle + \ldots\,,
\label{PsiHR}
\end{eqnarray}
where $\ldots$ stand for the non-resonant terms and
\begin{equation}
    c_{\mathcal{R}}^{MB} 
          = {{\cal V}^M_{B\mathcal{R}}
             \over Z_{\mathcal{R}}(W) (W-M_{\mathcal{R}})}\,.
\label{cMBR}
\end{equation}
We introduce a shorthand notation
$$
    g_{MB} \equiv g_\alpha 
       = \sqrt{\pi}\mathcal{N}_{MB}{\cal V}^{M}_{B\mathcal{R}}\,,
$$
which, at the resonance, is just the square root of the half 
width $\sqrt{\Gamma_{MB}/2}$ of the $MB$ channel.
Then
\begin{eqnarray}
    K_{\alpha\beta} &=& { g_\alpha g_\beta\over M_{\mathcal{R}}-W }
 + \ldots
  \equiv (g_1^2+g_2^2) {r_\alpha r_\beta\over (M_{\mathcal{R}}-W)}
 + \ldots
\nonumber \\
     r_i &=& {g_i\over\sqrt{g_1^2+g_2^2}}\,.
\end{eqnarray}
For simplicity we restrict ourselves to the case of only two channels.
The $K$ matrix can be cast in the form
\begin{equation}
    \mathbf{K} = 
        {(g_1^2+g_2^2)\over(M_{\mathcal{R}}-W)}
        \left|\matrix{r_1 & -r_2 \cr r_2 & r_1}\right|
        \left|\matrix{1 & 0 \cr 0 & 0}\right|
        \left|\matrix{r_1 & r_2 \cr -r_2 & r_1}\right|+ \ldots
\label{Kmatrix}
\end{equation}
{\em i.e.\/} the $K$ matrix is proportional to the projector operator.
It is then easy to derive the expression
\begin{eqnarray}
 &&  \left(\mathbf{1} + \mathbf{K}^2\right)^{-\h} =
 \nonumber \\ 
 &&  \left|\matrix{r_1 & -r_2 \cr r_2 & r_1}\right|
   \left|\matrix{{|M_{\mathcal{R}}-W|\over
           \sqrt{(M_{\mathcal{R}}-W)^2+(g_1^2+g_2^2)^2}} & 0 \cr
                   0  & 1  }\right|
        \left|\matrix{r_1 & r_2 \cr -r_2 & r_1}\right|+ \ldots
\nonumber \\
   &&= {|M_{\mathcal{R}}-W|\over
           \sqrt{(M_{\mathcal{R}}-W)^2+(g_1^2+g_2^2)^2}}
       \left|\matrix{r_1^2 & r_1r_2 \cr r_1r_2 & r_2^2}\right|
\nonumber \\ && \kern60pt
     + \left|\matrix{r_2^2 & -r_1r_2 \cr -r_1r_2 & r_1^2}\right|
     + \ldots \,.
\label{sqrt1pK2}
\end{eqnarray}
From (\ref{sol4K}) and (\ref{cMBR}) it follows that close to 
the resonance the state on  the RHS of (\ref{orthonorm}) can 
be put in the form
$$
   |\Psi^\beta\rangle = {K_{\alpha\beta}\over\sqrt{\pi} g_\alpha}
  |\widehat{\Psi}_{\mathcal{R}}\rangle\,.
$$
Evaluating the sum by using the expression (\ref{Kmatrix}) 
we notice that the second term in (\ref{sqrt1pK2}) vanishes 
\begin{eqnarray}
  |\widetilde{\Psi}^\alpha\rangle 
  &=& \sum_\beta  \left(\mathbf{1} 
    + \mathbf{K}^2\right)^{-\h}_{\beta\alpha}
   |\Psi^\beta\rangle
\nonumber \\
   &=& {g_\alpha\over\sqrt\pi\sqrt{(M_{\mathcal{R}}-W)^2+(g_1^2+g_2^2)^2}}
     |\widehat{\Psi}_{\mathcal{R}}\rangle + \ldots
\nonumber \\
  &=& 
   {1\over\sqrt{2\pi}}{\sqrt{\Gamma_{MB}}
 \over\sqrt{(M_{\mathcal{R}}-W)^2+{1\over4}\Gamma^2}}
      |\widehat{\Psi}_{\mathcal{R}}\rangle + \ldots\,,
\end{eqnarray}
where $\Gamma = \sum_{M'B'}\Gamma_{M'B'}$\,.
The factor in front of the resonance state enters in 
the calculation of the $K$-matrix elements corresponding
to the decay into this particular resonance, ${\cal R}$.
In fact, only the square of the factor appears
$$
    w_{MB}(M) = {1\over2\pi}{\Gamma_{MB}(M) \over 
                (M_{\mathcal{R}}-M)^2+{1\over4}\Gamma^2(M)}\,,
$$
where $M$ is now used for the invariant mass of the $MB$ system.
The weights $w_{MB}(M)$ are calculated in different partial
waves and then stored using a spline approximation.

The result can be readily generalized to the case of
three or more channels.


\section{\label{rhoN}The $\rho qq$ and $\sigma qq$ vertices}
The form of the $\rho$-meson coupling to the quarks in the 
Cloudy Bag Model has been discussed in \cite{Alvarez-Estrada83}; 
they suggested a pion-like coupling at the bag surface.
We therefore assume the Cloudy Bag Model type coupling
\begin{eqnarray}
   H_{\rho qq} &=& {\i\over2\tilde{f}_\rho}\int\d\vec{r}\,\delta(r-R)
   \sum_t\psi^\dagger\vec{\alpha}\tau_t\psi\vec{A}_t\,.
\label{Hrhoqq}
\end{eqnarray}
Here $\tilde{f}_\rho$ is analogous to $f_\pi$ in the pion-quark
interaction but there is no {\em a priori\/} reason to identify this
parameter with the $\rho$ decay constant,
$\tilde{f}_\rho  \approx 200$~MeV.

In order to derive the form of the interaction for a particular
partial wave it is most suitable to expand the $\rho$ field 
in the basis with good total angular momentum $J$,
its third component $M$, the orbital momentum $l$,
and the third component of the isospin $t$:
\begin{equation}
   \vec{A}_t = \sqrt{2\over\pi}\,\int{k\,\d k\over\sqrt{2\omega_k}}
   \sum_{JlM}\,j_l(kr) \vec{Y}_{JlM}(\hat{\vec{r}})   a_{JlMt}(k) 
   + \hbox{h.c.}
\label{AtJM}
\end{equation}
Here $\vec{Y}_{JlM}(\hat{\vec{r}})$ are the vector spherical 
harmonics and $a_{JlMt}(k)$ the meson annihilation operator.
It is related to the corresponding operator of the
plane-wave representation by
\begin{equation}
    a_{JlMt}(k) = 
        \i^l\, k \int \d\hat{\vec{k}} \sum_{\lambda m}
        Y_{lm}^*(\hat{\vec{k}})\CG{l}{m}{1}{\lambda }{J}{M}
        a_{\lambda t}(\vec{k})\,,
\label{aJlM}
\end{equation}
where $\lambda$ is the polarization ($\lambda=0,\pm1$) and 
$\CG{l}{m}{1}{\lambda }{J}{M}$ is the Clebsch-Gordan coefficient.

The $N\rho$ channel can be labeled, in addition to specifying 
the total angular momentum $J_{\mathrm{ch}}$ and $M_{\mathrm{ch}}$, 
by the relative angular momentum $l$ and either by the spin $S$ 
of the $\rho N$ system or by the total angular momentum $J_\rho$ of 
the $\rho$-meson.
The two basis states are related through
\begin{eqnarray}
 |SlJ_{\mathrm{ch}} M_{\mathrm{ch}}\rangle 
   &=&\sum_{J_\rho}\sqrt{2S+1}\sqrt{2J_\rho+1}\;
  W(lJ_{\mathrm{ch}}1\half;SJ_\rho) 
\nonumber\\ &&\times
  |J_\rho J_{\mathrm{ch}} M_{\mathrm{ch}} l\rangle \,,
\end{eqnarray}
where $ W(lJ_{\mathrm{ch}}1\half;SJ_\rho)$ is the Racah coefficient.

For the $\rho qq$ interaction involving the $s$-state
and the $p_{j=1/2}$-state quarks (SP) appearing in the negative
parity partial wave we obtain
\begin{equation}
    H^{SP}_{\rho\;Jl} = \int \d k\,
     V^{\rho{SP}}_{Jl}(k)
    \sum_{i=1}^3\sum_{Mt} \sigma_M(i)\tau_t(i)
      a_{1lMt}(k) + {\rm h.c}\,,
\end{equation}
with possible values $J=1$ and $l=0,2$:
\begin{eqnarray}
     V^{\rho{SP}}_{10}(k) &=& {1\over4\pi\tilde{f}_\rho}
     \sqrt{\omega_{p_{1/2}}\omega_s\over(\omega_{p_{1/2}}+1)(\omega_s-1)}\,
     {2\over3}\,
     {k^2\over\sqrt{\omega_k}}\,{j_0(kR)\over kR}\,,
\nonumber\\
     V^{\rho{SP}}_{12}(k) &=& {1\over4\pi\tilde{f}_\rho}
     \sqrt{\omega_{p_{1/2}}\omega_s\over(\omega_{p_{1/2}}+1)(\omega_s-1)}\,
     {\sqrt2\over3}\,
     {k^2\over\sqrt{\omega_k}}\,{j_2(kR)\over kR}\,.
\nonumber
\end{eqnarray}
Here $\sigma_M$ acts between the total angular momenta
of the quarks instead of their spins.
For the $\rho qq$ interaction between the $s$-state
and the $p_{j=3/2}$-state quarks (SA) we find
\begin{equation}
    H^{SA}_{\rho\;Jl} = V^{\rho{SA}}_{Jl}(k)\sum_{i=1}^3\sum_{Mt} 
     \Sigma_{JM}^{[\th\h]}(i)\tau_t(i)
     a_{JMt}(k) + {\rm h.c.}
\end{equation}
with $J=1,\; l=0,2$ and  $J=2,\; l=2$:
\begin{eqnarray}
     V^{\rho{SA}}_{10}(k) &=& {1\over4\pi\tilde{f}_\rho}
     \sqrt{\omega_{p_{3/2}}\omega_s\over(\omega_{p_{3/2}}-2)(\omega_s-1)}\,
     {1\over\sqrt3}\,
     {k^2\over\sqrt{\omega_k}}\,{j_0(kR)\over kR}\,,
\nonumber\\
     V^{\rho{SA}}_{12}(k) &=& -{1\over4\pi\tilde{f}_\rho}
     \sqrt{\omega_{p_{3/2}}\omega_s\over(\omega_{p_{3/2}}-2)(\omega_s-1)}\,
     {1\over\sqrt6}\,
     {k^2\over\sqrt{\omega_k}}\,{j_2(kR)\over kR}\,,
\nonumber\\
     V^{\rho{SA}}_{22}(k) &=& -{1\over4\pi\tilde{f}_\rho}
     \sqrt{\omega_{p_{3/2}}\omega_s\over(\omega_{p_{3/2}}-2)(\omega_s-1)}\,
     \sqrt{3\over2}\,
     {k^2\over\sqrt{\omega_k}}\,{j_2(kR)\over kR}\,,
\nonumber
\end{eqnarray}
where $\langle \thalf m_j|\Sigma_{JM}^{[\th\h]}|\half m_s\rangle
= \CG{\h}{m_s}{J}{M}{\th}{m_j}$.

In the case of the interaction involving the $s$-state 
quarks alone, only the $\rho$-mesons with $J=l=1$ contribute, 
{\em i.e.\/} only the transverse magnetic M1 multipole is present:
\begin{eqnarray}
   H_{\rho qq} &=& {1\over2\tilde{f}_\rho}
 \left({\omega_s\over\omega_s-1}\right)
{1 \over2\pi}\sqrt{2\over3}\,\int {\d k\,k^2\over\sqrt{\omega_k}}
       \,{j_1(kR)\over kR}
\nonumber\\ && \times
\sum_{i=1}^3\sum_{tM}\sigma_M(i)\tau_t(i)\left[
        a_{11Mt}(k) + \hbox{h.c.}\right.
\end{eqnarray}
By using  (\ref{aJlM}) this expression reduces to a more familiar form
\begin{eqnarray}
  H_{\rho qq} &=& {\i\over2\tilde{f}_\rho}
  \left({\omega_s\over\omega_s-1}\right)
  {1 \over3\sqrt{2\pi}^3}\int {\d\vec{k}\over\sqrt{2\omega_k}}
  \,{3j_1(kR)\over kR}\,
\nonumber\\ && \times
\sum_{i=1}^3\sum_{t\lambda }\tau_t(i)(\vec{\sigma}(i)\times\vec{k})\cdot
   \vec{\varepsilon}_\lambda  \,a_{\lambda t}(\vec{k}) + {\rm h.c.}\,,
\nonumber\\
\end{eqnarray}
which should be compared to the corresponding result in the pion case
\cite{CBM}
\begin{eqnarray}
  H_{\pi qq} &=&{\i\over2f_\pi}\left({\omega_S\over\omega_S-1}\right)
  {1 \over3\sqrt{2\pi}^3}\int {\d\vec{k}\over\sqrt{2\omega_k}}
  \,{3j_1(kR)\over kR}
\nonumber\\ && \times
\sum_{i=1}^3\sum_{t}\tau_t(i)\,\vec{\sigma}(i)\cdot\vec{k}
   \,a_{t}(\vec{k}) + {\rm h.c.}
\end{eqnarray}
We are now able to relate $\tilde{f}_\rho$,  which determines 
the strength of the interaction in the Cloudy Bag Model,
to the corresponding parameter in the pion case, $f_\pi$,
in terms of the meson masses and the coupling
constants $f_{\pi NN}$ and $f_{\rho NN}$.
In either case, relating the expectation value of the quark 
operators $\sum_{i=1}^3\tau(i)\sigma(i)$ in the nucleon to
the corresponding operators acting on the nucleon isospin 
and spin, the same factor of 5/3 appears.
Hence
$$
    \tilde{f}_\rho = {m_\rho f_{\pi NN}\over m_\pi f_{\rho NN}}\, f_\pi\,.
$$
Here $f_{\rho NN}$ reads \cite{Matsuyama07}
$$
    {f_{\rho NN}\over m_\rho} = {g_{\rho NN}(1+k_\rho)\over 4m_N}\,.
$$
For typical values of $g_{\rho NN}$ and $k_\rho$  
(see {\em e.g.} \cite{Matsuyama07,ronchen13}) one obtains
$
    \tilde{f}_\rho \approx (1.5 \div 2)\; f_\pi\,.
$

For the weight function multiplying the $\rho$-meson vertices
we assume a Breit-Wigner form modified by an energy-dependent
correction involving the $\rho$ range parameter \cite{PDG}.

As stressed in ref.~\cite{Alvarez-Estrada83}, the number of $\rho$ mesons 
is extremely small in the nucleon which means that the rho-meson loops 
contribute little to the self energy and the vertex renormalizations.  
These contributions have therefore not been taken into account in our 
calculation.

The $\sigma$-quark interaction cannot be derived in an analogous
way since it disappears at the bag surface.
We therefore use a purely phenomenological approach as in \cite{EPJ2008}.
For the $s$-wave we have assumed
$$
   V^\sigma_0(k,\mu) 
   = V^\sigma_0(k)\,w_\sigma(\mu)    \,,
\qquad
    V^\sigma_0(k) 
   = G_\sigma{k\over\sqrt{2\omega_{k}}}\,.
$$
Here $\omega_{k}^2 = k^2 + \mu^2$, $\mu$ is the invariant mass of the 
two-pion system and $w_\sigma(\mu)$ is a Breit-Wigner weight function 
centered around $m_\sigma=450$~MeV with the width $\Gamma_\sigma=550$~MeV; 
$G_\sigma$ is a free parameter determined in the $N(1440)$ decay.
For the $p$-wave we assume
$$
    V^\sigma_{1m}(k) = V^\sigma_0(k)\, {kR\over3}\,\sum_{i=1}^3
  \left[{1\over\sqrt3}\,\sigma_m(i) + \Sigma_{1m}^{\th\h}(i)\right]\,,
$$
where $\sigma_m$ involves transition to the $p_\h$ quark state and  
$\Sigma_{1m}^{\th\h}$ (defined above) to the $p_\th$ state.

\end{document}